\documentclass[pra,twocolumn,nofootinbib,floatfix,10pt]{revtex4-2}

\usepackage{amsmath}
\usepackage{amssymb}
\usepackage{wasysym}
\usepackage{graphicx}
\usepackage{color,soul}
\usepackage{physics}
\usepackage{siunitx}
\usepackage{dsfont}
\usepackage{float}
\usepackage[english]{babel}
\usepackage{blindtext}
\usepackage[english,nomargin,inline,marginclue,draft]{fixme}
\pdfpageheight\paperheight
\pdfpagewidth\paperwidth

\usepackage[colorlinks,linkcolor=blue,anchorcolor=blue,citecolor=blue,urlcolor=blue]{hyperref}

\fxusetheme{colorsig}
\FXRegisterAuthor{cg}{acg}{CG}  
\FXRegisterAuthor{th}{ath}{\color{blue}TH}  
\FXRegisterAuthor{ib}{aib}{\color{red}IB} 
\FXRegisterAuthor{sh}{ash}{\color{cyan}SH} 
\FXRegisterAuthor{db}{adb}{\color{green}DB} 
\FXRegisterAuthor{ps}{aps}{PS}
\makeatletter
\renewcommand*\FXLayoutInline[3]{%
  {\@fxuseface{inline}\ignorespaces{\color{fx#1}[#3: #2]}}}
\makeatother

\long\def\symbolfootnote[#1]#2{\begingroup%
\def\thefootnote{\fnsymbol{footnote}}\footnotetext[#1]{#2}\endgroup}

\def\nobreakbefore{%
  \relax\ifvmode\else
    \ifhmode
      \ifdim\lastskip > 0pt\relax
        \unskip\nobreakspace
      \else 
        \nobreakspace
      \fi
    \fi
  \fi
}
\let\oldcite\cite
\renewcommand\cite{\nobreakbefore\oldcite}

\begin{document}
\title{Quantum enhanced metrology based on flipping trajectory of cold Rydberg gases}

\author{Ya-Jun Wang$^{1,2,\textcolor{blue}{\star}}$}
\author{Jun Zhang$^{1,2,\textcolor{blue}{\star}}$}
\author{Zheng-Yuan Zhang$^{1,2,\textcolor{blue}{\star}}$}
\author{Shi-Yao Shao$^{1,2}$}
\author{Qing Li$^{1,2}$}
\author{Han-Chao Chen$^{1,2}$}
\author{Yu Ma$^{1,2}$}
\author{Tian-Yu Han$^{1,2}$}
\author{Qi-Feng Wang$^{1,2}$}
\author{Jia-Dou Nan$^{1,2}$}
\author{Yi-Ming Yin$^{1,2}$}
\author{Dong-Yang Zhu$^{1,2}$}
\author{Qiao-Qiao Fang$^{1,2}$}
\author{Chao Yu$^{1,2}$}
\author{Xin Liu$^{1,2}$}
\author{Guang-Can Guo$^{1,2}$}
\author{Bang Liu$^{1,2,\textcolor{blue}{\ddagger}}$}
\author{Li-Hua Zhang$^{1,2,\textcolor{blue}{\S}}$}
\author{Dong-Sheng Ding$^{1,2,\textcolor{blue}{\dagger}}$}
\author{Bao-Sen Shi$^{1,2}$}

\affiliation{$^1$Laboratory of Quantum Information, University of Science and Technology of China, Hefei, Anhui 230026, China.}
\affiliation{$^2$Anhui Province Key Laboratory of Quantum Network, University of Science and Technology of China, Hefei 230026, China.}

\date{\today}
\symbolfootnote[1]{Y.J.W, J.Z, Z.Y.Z contribute equally to this work.}
\symbolfootnote[3]{lb2016wu@ustc.edu.cn}
\symbolfootnote[4]{zlhphys@ustc.edu.cn}
\symbolfootnote[2]{dds@ustc.edu.cn}

\maketitle
\textbf{The dynamical trajectory of a dissipative Rydberg many-body system could be flipped under a microwave field driving, displaying an enhanced sensitivity. This is because the intersection of the folded hysteresis trajectories exhibits a sharp peak near the phase transition, amplifying the response to small changes in the microwave field. Here, we demonstrate an experiment of enhanced metrology through flipping the hysteresis trajectory in a cold atomic system, displaying an approach to improve sensitivity near the gap-closing points. By measuring the intersection points of hysteresis trajectories versus Rabi frequency of the microwave field, we quantify the equivalent sensitivity to be 1.6(5) $\rm{nV}cm^{–1}Hz^{–1/2}$. The measurement is also dependent on the interaction time, optical depth and principal quantum number since the long-range interaction between Rydberg atoms could dramatically change the shape of hysteresis trajectories. The reported results suggest that flipping trajectory features in cold Rydberg many-body systems could advance sensing and metrology applications.}

Quantum criticality has been recognized as a key resource for achieving enhanced sensitivity. In equilibrium systems, various critical phenomena including second-order~\cite{jin2005first,sarkar2022free}, localization \cite{li2020critical,zhang2022review}, and topological phase transitions \cite{lee2019topological,de2019observation,porta2020topological} have been utilized for sensing applications. For non-equilibrium systems, quantum-enhanced sensitivity has been observed in dissipative phase transitions \cite{perez2018glassy,hao2023topological,wu2024dissipative}, Floquet systems \cite{jiang2022floquet}, and time crystal phases \cite{liu2024higher,liu2025bifurcation,o2025quantum}. A common feature across these critical systems that enables quantum-enhanced sensitivity is the closing of energy gaps. In non-Hermitian systems, this gap-closing effect modifies the system's fundamental properties, leading to new physical phenomena such as symmetry breaking \cite{lourencco2022non,chen2023continuous},  Yang-Lee edge singularity \cite{shen2023proposal,gao2024experimental}, and the appearance of exceptional points \cite{wang2022non,li2023synergetic}. These properties are extensively investigated in quantum systems like condensed matter \cite{laflorencie2016quantum,ma2019topological}, optics \cite{barik2018topological,parto2020non}, circuits \cite{roushan2014observation,imhof2018topolectrical}, and cold atom systems \cite{yoshida2018reduction,li2020topological}. Recently, the critical behavior of many-body systems has emerged as a promising framework for enhanced sensing \cite{montenegro2021global,ilias2022criticality,montenegro2024quantum}, particularly in the form of near-critical quantum-enhanced sensing. Proximity to the exceptional point, the system's eigenvalues exhibit a heightened sensitivity to small parameter changes, thus improving the precision of sensing operations \cite{lau2018fundamental,budich2020non,li2023synergetic}. Hence, the gap-closing has been demonstrated to be a prime contender for the ultimate origin of quantum-enhanced sensing \cite{sarkar2022free,koch2022quantum,sarkar2024critical,sarkar2025exponentially}.

Rydberg atoms, characterized by their dipole moments \cite{saffman2010quantum,adams2019rydberg,browaeys2020many}, exhibit unique strong interactions and specific energy level configurations that make them promising candidates for studying non-equilibrium dynamics \cite{lee2012collective,carr2013nonequilibrium,schempp2014full,marcuzzi2014universal,lesanovsky2014out,urvoy2015strongly, Signatures2020Helmrich,ding2019Phase,Wintermantel2020Cellular, klocke2021hydrodynamic,gambetta2019, Wadenpfuhl2023Synchronization, ding2023ergodicity,liu2024microwave,ma2024microwave}, topology \cite{samajdar2021quantum,kanungo2022realizing,li2021symmetry,li2015exotic,verresen2021prediction} and precise measurement \cite{sedlacek2012microwave,jing2020atomic,ding2022enhanced,liuHighly,zhang2022rydberg,zhang2024ultra,liu2023electric,zhang2024rydberg}, and etc. Rydberg atomic systems also exhibit rich hysteretic trajectories that arise from interaction-induced non-Hermitian properties~\cite{delplace2021symmetry,zhang2025exceptional}, with a physical mechanism  distinct from that of bistability~\cite{carr2013nonequilibrium,zhang2025microwave}. Building on our previous discovery of interaction-induced exceptional points and hysteretic phenomena in cold Rydberg gases, we further investigate quantum sensing based on exceptional points. The interplay between this unique non-Hermitian dynamics of Rydberg atom systems and their complex critical properties opens exciting tools for applications in sensing microwave fields. The complex energy spectrum of non-Hermitian systems is highly sensitive to changes in boundary conditions; the subtle changes could lead to the transition of distinct phases; this results in a robust and exponentially enhanced sensitivity \cite{budich2020non}.

\begin{figure*}
\centering
\includegraphics[width=1\linewidth]{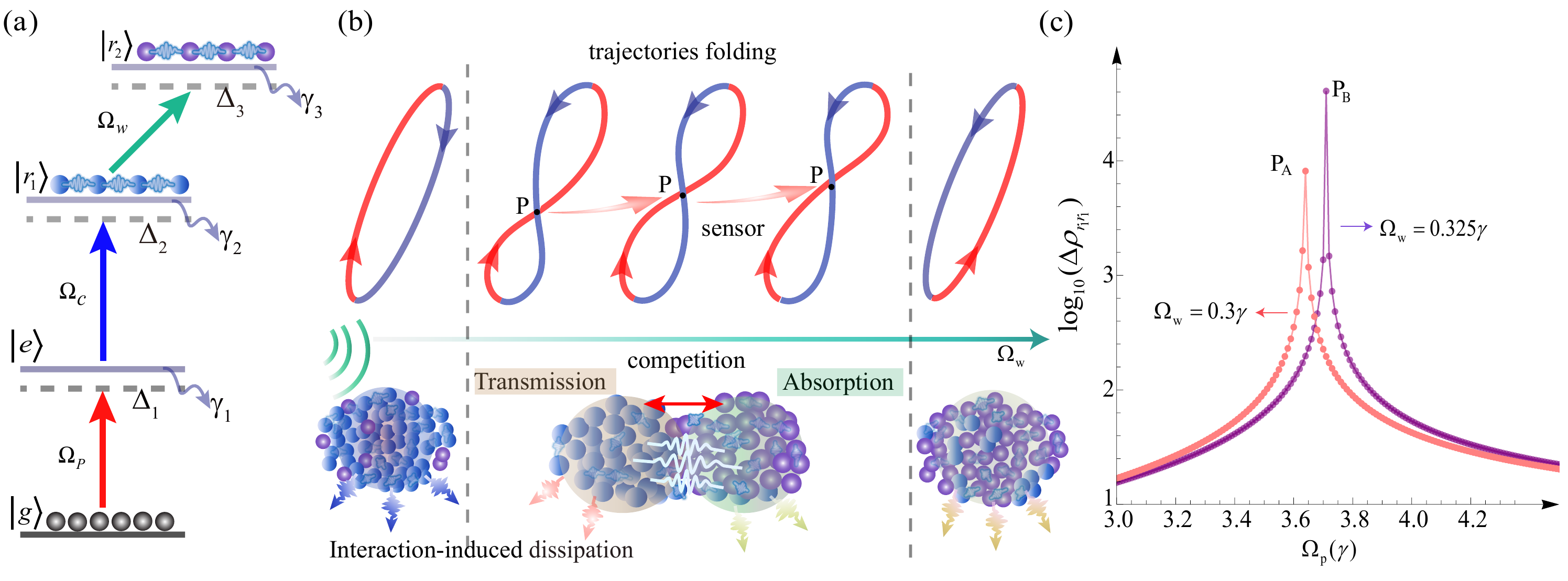}
\caption{\textbf{Diagram of enhanced metrology based on flipping dynamical trajectories.} (a) The energy level diagram consists of a ground state \(\ket{g}\), an intermediate excited state \(\ket{e}\), a Rydberg state \(\ket{r_1}\), and another Rydberg state \(\ket{r_2}\). The probe field (red), with a Rabi frequency \(\rm{\Omega_{p}}\) and a detuning \(\Delta_{1}\), couples the transition between the ground state \(\ket{g}\) and the excited state \(\ket{e}\). Meanwhile, the coupling field (blue) with a Rabi frequency \(\rm{\Omega_{c}}\) and a detuning \(\Delta_{2}\) drives the transition from \(\ket{e}\) to \(\ket{r_1}\). A microwave field (green) coupling the transition between \(\ket{r_1}\) and \(\ket{r_2}\) is used to demonstrate the enhanced measurement; it has a Rabi frequency \(\rm{\Omega}_{w}\) and a detuning \(\Delta_{3}\). (b) Under microwave field driving, the system's response trajectories go through two stages with opposite rotation directions, accompanied by an intermediate stage of a folded trajectory state (see upper panel). The folding intersection point (P) exhibits enhanced sensitivity to the microwave field, where the interplay between the absorption and transmission of the interacting Rydberg atoms contributes to precise measurements (see lower panel). (c) The difference $\Delta\rho_{r_1r_1}$ in transmission between the upward and downward scanning trajectories for $\rm{\Omega_{w}}=0.3\gamma$ and $\rm{\Omega_{w}}=0.325\gamma$. }
\label{Fig.1}
\end{figure*}

Typically, traditional methods often involve analyzing how an external field affects the spectral properties of a system—from the energy levels of atoms or molecules to the absorption or transmission spectra \cite{sedlacek2012microwave,ding2022enhanced,liu2023electric}. In this case, the sensitivity is dependent on the bandwidth or slope of the spectrum. When a microwave field is applied to the interacting Rydberg atoms, it induces a change in the energy spectrum, leading to the closure of energy gaps between different eigenstates. This `gap-closing' effect results in the emergence of `folded' hysteresis trajectories or the transition between distinct phases. In this scenario, the non-Hermitian properties experience a dramatic change versus the microwave field, thus can be regarded as a resource to demonstrate enhanced sensing. In this work, we demonstrate a precise measurement of the microwave field based on flipping trajectories of cold Rydberg gases. Under the external microwave fields driving, the hysteresis trajectory of the atoms' transmission exhibits distinct folding patterns with increasing and decreasing the probe Rabi frequency, leading to the formation of microwave field-dependent intersection points. By analyzing the shifts in the intersection position in response to the external microwave field, we can accurately measure the magnitude of the external field with high sensitivity. 

\subsection*{Physical model}
To demonstrate enhanced metrology, we firstly build a model with a configuration of an interacting four-level atom, as given in Fig.~\ref{Fig.1}(a).  In the rotating frame, the effective Hamiltonian of the system is written as 
\begin{align}
H=&-\sum\limits_j [2{\Delta _1}\sigma _{ee}^j + (2\Delta_{r_1}+iV_{\rm{eff}})\sigma _{r_{1}r_{1}}^j + {2\Delta _{r_2}}\sigma _{r_{2}r_{2}}^j] \nonumber\\ 
&+\sum\limits_j [({{\rm{\Omega} _p}\sigma _{ge}^j + {\rm{\Omega _c}}\sigma _{er_1}^j + {\rm{\Omega _{w}}}\sigma _{r_{1}r_{2}}^j) + H.c.}],
\end{align} 
where $\sigma _{ab}^j = \left| {{a_j}} \right\rangle \left\langle {{b_j}} \right|(a,b = g,e,r_1,r_2)$, $\Delta_{r_1}=\Delta_1+\Delta_2$, $\Delta_{r_2}=\Delta_1+\Delta_2+\Delta_3$, $H.c.$ denotes Hermitian conjugate. The dissipative effect induced by the interaction between Rydberg atoms can be effectively described by the non-Hermitian term $V_{\rm{eff}} = V \rho_{r_1r_1}$ \cite{zhang2025exceptional}, where $\rho_{r_1r_1}$ is the population of the $\ket{r_1}$ state and $V$ denotes the interaction strength. Under the lasers' excitations, the system's response displays an intermediate state [Fig.~\ref{Fig.1}(b)], at which the probe transmission trajectories are folded and result in an intersection point P. The transmission spectra at intersection points exhibit resolvable splitting even for minute microwave detunings (The intersection points $\rm{P_A}$ and $\rm{P_B}$ in Fig.~\ref{Fig.1}(c)), enabling high sensitivity microwave field detection.

\begin{figure*}
\centering
\includegraphics[width=1\linewidth]{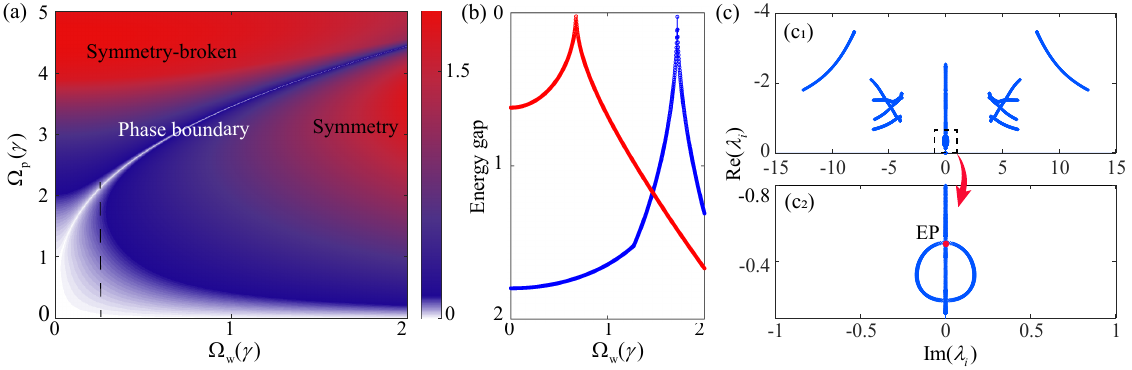}
\caption{\textbf{Energy gap phase diagram.} (a) The energy gap is obtained by numerically solving the modulus of $E_2-E_3$, where the legend represents the magnitude of the energy gap. The white curve with a zero energy gap separates the particle-hole symmetric phase from the symmetry-broken phase. The position of the black dashed line is $\rm{\Omega_{w}}=0.3\gamma$. The parameters setting are $\rm{\Omega_{c}}=6\gamma$ and $V_{\rm{eff}}= V \rho_{r_1r_1}$ $(\rho_{r_1r_1}\propto\rm{\Omega_{p}}^2)$. (b) The energy gap at $\rm{\Omega_{p}}=3\gamma$ (red solid line) and $\rm{\Omega_{p}}=4.2\gamma$ (blue solid line). (c) The Liouvillian eigenspectrum complex planes of the system for $\{\rm{\Omega_{w}},\rm{\Omega_{c}},V_{\rm{eff}},\gamma_1,\gamma_2,\gamma_3\} =\{0.3\gamma, 6\gamma, 0.6\gamma,\gamma,2\gamma,0.5\gamma\}$ in ($\rm{c}_1$) and the enlarged view in ($\rm{c}_2$), where the Liouvillian eigenvalues evolve as the parameter $\Omega_{p}$ varies within the range [0 - 5$\gamma$].}
\label{Fig.2}
\end{figure*}
Figure.~\ref{Fig.1}(b) illustrates the physical mechanism underlying the enhanced sensing in the Rydberg atomic system. As $\rm{\Omega_{w}}$ varies, the complex interaction of the many-body system induces a folding of the hysteresis loop's rotation direction from clockwise to counterclockwise. During the trajectories reversal, the system exhibits an intermediate process of $\rm{\Omega_{p}}$ up-scanning trajectories folding with $\rm{\Omega_{p}}$ down-scanning trajectories. This phenomenon arises from the `gap closing' effect in the complex energy plane, which induces spontaneous breaking of particle-hole symmetry. Within this regime, the system undergoes a competition between absorption and transmission, resulting in the intersection of hysteresis loops. The intersection point of the trajectories is regarded as a highly resolvable observable that aligns with the microwave Rabi frequency $\rm{\Omega_{w}}$, allowing us to demonstrate enhanced sensing. The transition from non-intersecting to intersecting trajectories represents a critical behavior, which significantly enhances the measurement sensitivity of the system. 

Theoretically, we study non-Hermitian hysteresis loops by the Lindblad master equation $\dot{\rho}=-i[H, \rho]+L[\rho]:=\mathcal{L}\rho$, where the operator $L[\rho]=\sum_j\left(L_j \rho L_j^{\dagger}-\frac{1}{2}\left\{L_j^{\dagger} L_j, \rho\right\}\right)$ describes the decay of the system, and $\mathcal{L}$ is a Liouvillian superoperator, see more details in methods section. Here, we consider the decay $L[\rho] = \sqrt{\gamma_1}\ket{g} \bra{e}+\sqrt{\gamma_2}\ket{e} \bra{r_1}+\sqrt{\gamma_3}\ket{r_1} \bra{r_2}$. Then, we obtain the time-dependent evolution relationship of $\rho_{r_1r_1}$ dependent on $\rm{\rm{\Omega_{p}}}$ at different $\rm{\Omega_{w}}$, as depicted in Fig.~\ref{Fig.1}(a). The $\rho_{r_1r_1}$ values for the upward and downward scans of $\rm{\Omega_{p}}$ differ due to time dependence, and the corresponding Lindblad equations are as follows:
 \begin{align} 
\rho_{r_1r_1}(\rm{\Omega_{w}},{\rm{\Omega_{p}}}_ \uparrow )&=\int\limits_0^{t'}{\left\{-i[\hat{H},\rho_{r_1r_1}]+ \textcolor{blue}{L}[\rho_{r_1r_1}]\right\}dt}\\
\rho_{r_1r_1}(\rm{\Omega_{w}},{\rm{\Omega_{p}}}_ \downarrow )&=\int\limits_0^{t'}{\left\{-i[\hat{H},\rho_{r_1r_1}]+ \textcolor{blue}{L}[\rho_{r_1r_1}]\right\}dt}  
\end{align}
In order to obtain the intersection of trajectories, we calculate the difference $\Delta\rho_{r_1r_1}$ in transmission between the upward and downward scanned trajectories, which can be expressed as
\begin{align}
\Delta\rho_{r_1r_1}=\left|\rho_{r_1r_1}(\rm{\Omega_{w}},{\rm{\Omega_{p}}}_ \uparrow )-\rho_{r_1r_1}(\rm{\Omega_{w}},{\rm{\Omega_{p}}}_ \downarrow )\right|.
\end{align}
In the trajectories folding region, apart from the starting point and the ending point, the intersection point of two trajectories also features $\Delta\rho_{r_1r_1}=0$. Here, we ignore the starting point and the ending point, and under diverse microwave $\rm{\Omega_{w}}$ conditions, the corresponding Rabi frequency of the probe field can be identified at $\Delta\rho_{r_1r_1}=0$, as shown in Fig.~\ref{Fig.1}(c).

At zero energy detuning ($\Delta_1=\Delta_{r_1}=\Delta_{r_2}=0$), this non-Hermitian Hamiltonian exhibits particle-hole symmetry \cite{yuce2018stable}, satisfying the relation 
\begin{equation}
\mathcal CH^*{\mathcal C^\dagger } = - H,
\end{equation}
more detailed discussions in supplementary materials. By diagonalizing the Hamiltonian $\hat{H}$, we can obtain the energy band $E_1$,$E_2$,$E_3$ and $E_4$. The closure of the $E_2$ and $E_3$ energy gaps leads to the breaking of particle-hole symmetry in the system, resulting in symmetric and symmetry-broken phases. 

In Fig.~\ref{Fig.2}(a), we plotted the phase diagram of the energy gap between $E_2$ and $E_3$ in the parameter plane of $\rm{\Omega_{w}}$ and $\rm{\Omega_{p}}$. The particle-hole symmetric region ($E_2=-{E_3}^*$) and the symmetry-broken region (${\rm{Re}}(E_{2})=-{\rm{Re}}(E_{3})$ and ${\rm{Im}}(E_{2}) \ne {\rm{Im}}(E_{3})$) are separated by the gap-closing points (or exceptional points), in which the energy gap approaches zero along the white line (${\rm{Re}}(E_{2})-{\rm{Re}}(E_{3})=0$ and ${\rm{Im}}(E_{2})-{\rm{Im}}(E_{3})=0$). Figure~\ref{Fig.2}(b) presents a cross-sectional view of the phase diagram for $\rm{\Omega_{p}}=3\gamma$ and $\rm{\Omega_{p}}=4.2 \gamma$. We also see a phase boundary line in the phase diagram, where the exceptional point (EP) exhibit a non-linear dependence on external parameters (e.g., electric field $\rm{\Omega_{w}}$ and probe field $\rm{\Omega_{p}}$). This nonlinear response dramatically lowers the measurement limit for perturbations, theoretically enabling sensitivity beyond that of traditional sensors \cite{hokmabadi2019non}. It is different from the enhanced characteristics of complex frequency splitting at EPs~\cite{chen2017exceptional,lau2018fundamental}.

\begin{figure*}
\centering
\includegraphics[width=0.96\linewidth]{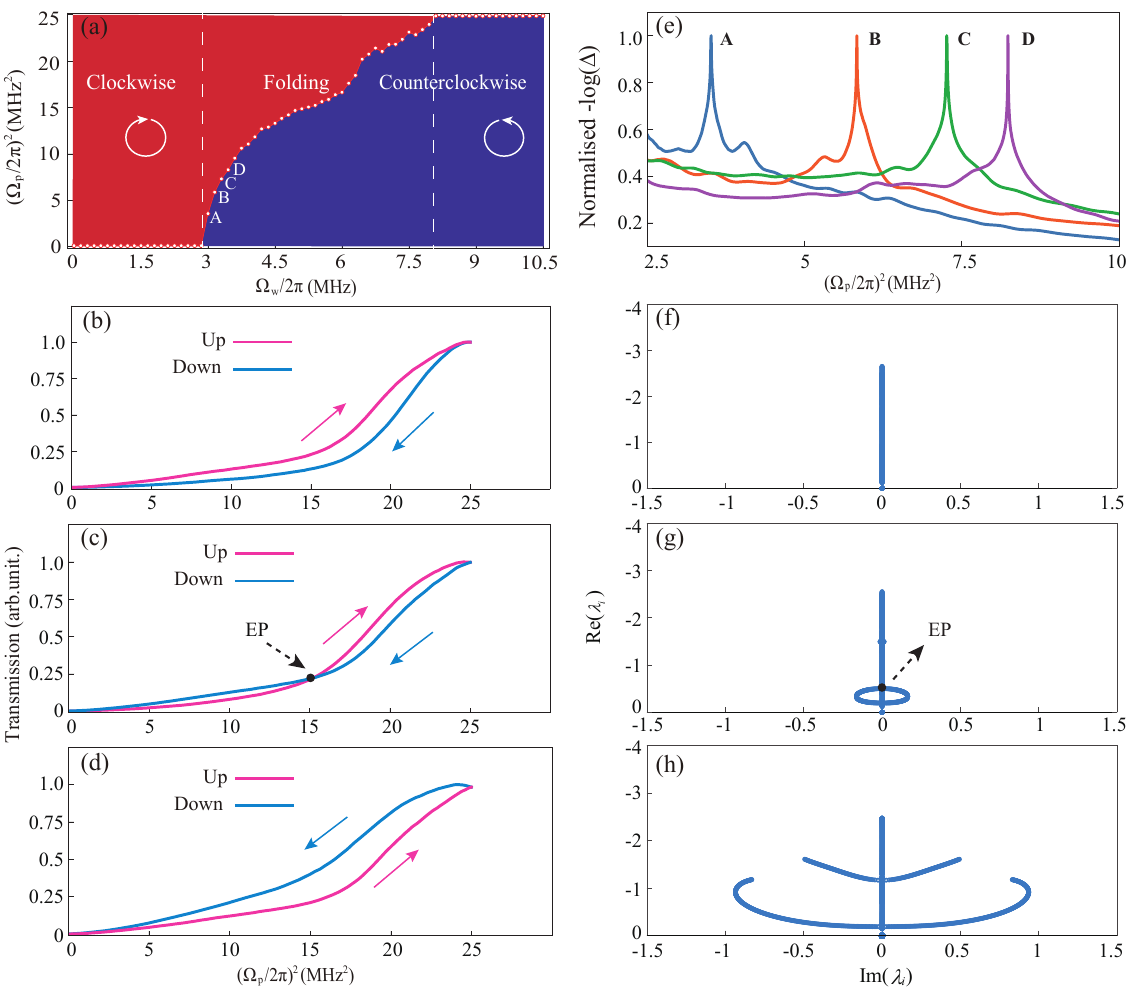}
\caption{\textbf{Phase diagram and hysteresis trajectories.} (a) The measured phase diagram of hysteresis trajectories. The red and blue regions indicate $\rm{T_{up}>T_{down}}$ and $\rm{T_{up} \leq T_{down}}$, respectively. (b)-(d) Evolution of the hysteresis trajectories under varying $\rm{\Omega}_{\rm{w}}$, where $\rm{\Omega}_{\rm{w}}$ are $2\pi\times $1.2 MHz in (b),  $2\pi\times $4.8 MHz in (c), and  $2\pi\times $9 MHz in (d). (e) The logarithm of the measured trajectories difference $-\rm{log(\Delta T})$, where blue, orange, green and purple correspond to the data at point A, B, C, and D in (a), respectively. (f)-(h) The Liouvillian eigenspectrum complex planes of the system are plotted as a function of $\rm{\rm{\Omega_{p}}}$ with $\rm{\Omega_{c}},V_{\rm{eff}},\gamma_1,\gamma_2,\gamma_3\} =\{4\gamma, 0.6\gamma,\gamma,2\gamma,0.5\gamma\}$  in cases of (f) $\rm{\Omega_{w}}= 0$, (g) $\rm{\Omega_{w}}$ =$0.3 \gamma$ and (h) $\rm{\Omega_{w}}$ = $\gamma$, where the parameter $\Omega_{p}$ varies within the range [0 - 5$\gamma$]. In (g), the curves intersect at an exceptional point (EP), marked by the shaded arrow, where the eigenvalues coalesce.}
\label{Fig3}
\end{figure*}

Similarly, to investigate the emergence of Liouvillian exceptional points during the dynamical evolution of the system, the Liouvillian eigenspectrum is obtained by solving $\mathcal{L}\hat{\rho}_i = \lambda_i \hat{\rho}_i$, where $\lambda_i$ and $\hat{\rho}_i$ denote the Liouvillian eigenvalues and eigenstates, respectively. Since there is one less $i$ factor compared with Schrodinger equation, the real parts of the eigenvalue of $\mathcal{L}$ and the imaginary parts of the eigenvalue of $H$ should be considered as mode gain or loss. The complex-plane distribution of the Liouvillian eigenspectrum is shown in Fig.~\ref{Fig.2}(c).  As expected, the enlarged view in Fig.~\ref{Fig.2}($\rm{c}_2$) demonstrates the emergence of exceptional point (EP) resulting from the coalescence of eigenvalues and eigenstates. Moreover, it reveals a remarkable transition in the characteristics of spontaneous symmetry breaking as the system traverses a gap-closing point. The real parts of these eigenvalues reflect the Rydberg interaction-induced dissipation and we are more concerned about the low-frequency part (The imaginary parts of the Liouvillian eigenvalues are close to zero in the system). These distinct eigenvalue configurations give rise to fundamentally different evolution trajectories, including characteristic trajectory folding phenomena, as further analyzed in the supplementary materials.

\begin{figure*}
\centering
\includegraphics[width=1\linewidth]{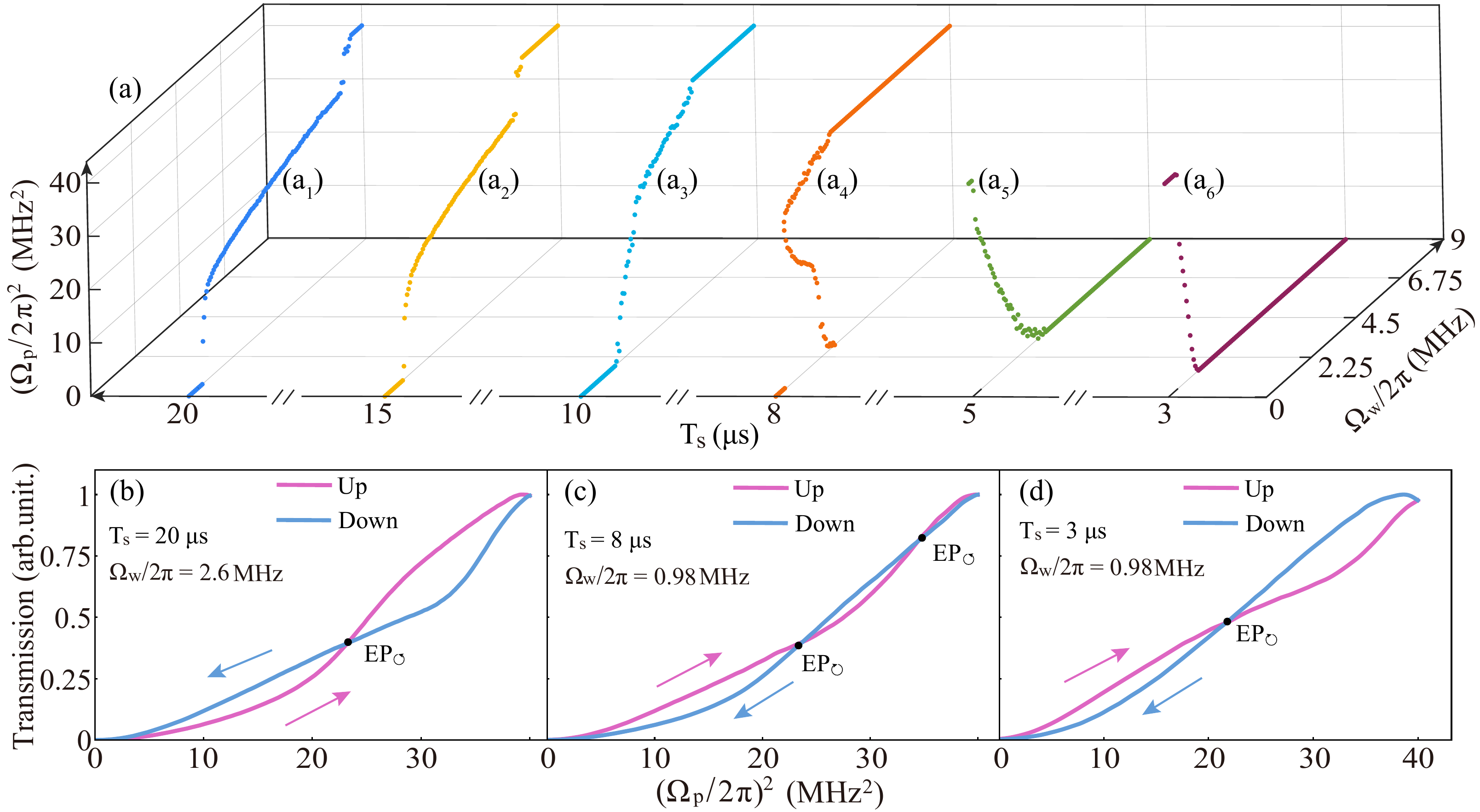}
\caption{\textbf{Measured phase boundaries versus scanning times.} (a) The phase boundaries measured at scanning times of 20 $\mu s$, 15 $\mu s$, 10 $\mu s$, 8 $\mu s$, 5 $\mu s $, and 3 $\mu s$ respectively. (b)-(d) The hysteresis trajectories measured as a function of $\rm{\Omega}_{\rm{p}}^2$ and $\rm{\Omega}_{\rm{w}}$ at scanning times of 20 $\mu s$, 8 $\mu s$, and 3 $\mu s$ respectively. The intersection points with the trajectory counterclockwise evolution marked by the red and blue arrows are defined as $\mathrm{EP}_{\circlearrowleft}$, and the case of the conversed evolution corresponds to $\mathrm{EP}_{\circlearrowright}$.}
\label{Fig4}
\end{figure*}

\section*{Results}
\subsection*{Enhancing sensitivity via trajectories folding}
In our system, we conduct experiments with a cold ensemble of \(^{85}\)Rb atoms utilizing electromagnetically induced transparency (EIT), see more details in methods section. The system's dynamic evolution is monitored by measuring probe transmission. We modify the probe intensity using a triangle waveform, generated by an acousto-optic modulator with a sweep period of \(\rm{T_s}\), and record the transmission trajectories during both the Up and Down processes. Due to the interactions between Rydberg atoms, the interactions-induced dissipation plays a vital role in the system's dynamic evolution. The increase and decrease of Rydberg atoms [by scanning $(\rm{\Omega_{p}}/2\pi)^2$] result in asymmetrical responses to probe transmission, generating a hysteresis loop \cite{zhang2025exceptional}.  

The measured hysteresis trajectories as a function of \(\rm{\Omega_{w}}\) enable us to map the full dynamics in the parameter space of \(\rm{(\Omega_{p}/2 \pi)^2}\) and \(\rm{\Omega_{w}/2 \pi}\). By recording the intersection points of the Up and Down processes, we obtain the phase diagram [Fig.~\ref{Fig3}(a)]. There are two phases with different color regions [$\rm{T_{up}>T_{\rm{down}}}$ (red) and $\rm{T_{up} \leq T_{\rm{down}}}$ (blue)], which are separated by the phase boundaries (the white intersection dots in Fig.~\ref{Fig3}(a)). The phase diagram includes three distinct hysteresis trajectories: clockwise rotating, folding, and counterclockwise rotating trajectories. Figures.~\ref{Fig3}(b)-(d) display the measured hysteresis trajectories under the distinct microwave field driving.  
\begin{figure*}
\centering
\includegraphics[width=1\linewidth]{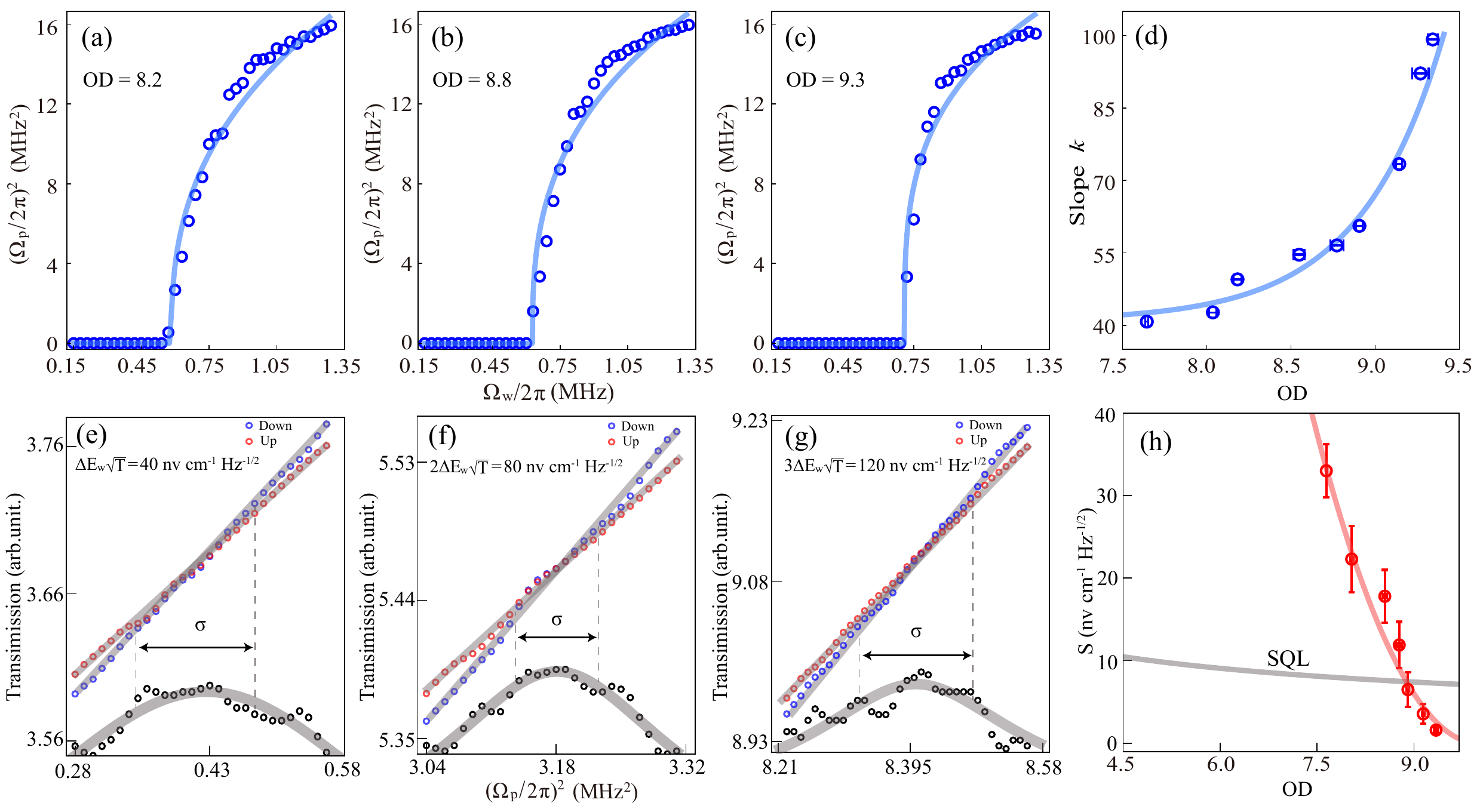}
\caption{\textbf{Measured phase boundaries and sensitivities against optical density (OD).} (a)-(c) The phase boundaries measured at OD of 8.2, 8.8, and 9.3 respectively. 
The blue line represents the fitted curve using a piecewise function: y = 0 for $x < \rm{\Omega_{wc}}$, and y = $p(x - \rm{\Omega_{wc}})^q$ for $x \geq \rm{\Omega_{wc}}$. Here, $\rm{\Omega_{wc}}$ represents the critical point of microwave field Rabi frequency, and the exponent $q$ takes values of 0.39, 0.35, and 0.28 for (a), (b), and (c), respectively. (d) The slopes of the phase boundaries at different ODs. The blue line fitted the phase boundaries' slope from the function y=$ae^{b(x-c)}+d$, where the parameters are $(a,~b,~c,~d)=(0.00072,~2,~3.7,~38)$. (e)-(g) The measured Up and Down transmission signals near the intersection point at OD=9.3(4) using another dataset with 1-3 step sizes, where the optical intensity fluctuations are represented by the gray shadows. The overlapping area of the gray shadows indicates the positional uncertainty range of intersection points, quantified by $\sigma$. The bottom inset displays the the differential transmission spectrum of the Up and Down processes with a fitted, provided as a visual guide. The gray curve is fitting curve with the form of a Lorentz function. The three panels correspond to measurements taken at microwave field strengths of 
$\rm{E_{wc}} + \Delta E_w$, $\rm{E_{wc}} + 2\Delta E_w$, and $\rm{E_{wc}} + 3\Delta E_w$, respectively, where $\rm{E_{wc}} = (\hbar\Omega_{\mathrm{wc}} \lambda_p )/  (\mu_{\mathrm{w}}  \lambda_c )$ and $\rm{\Delta E_w}$ denotes the step size of the applied microwave field. (h) The sensitivity as a function of OD, and the error bars represent the statistical analysis of the fluctuation in $\sigma$. The red curve is the fitting function y=$a{(x-b)^2}$ with parameters $(a,~b)=(6,~10)$, and the gray curve represents the standard quantum limit.}
\label{Fig5}
\end{figure*}

We also tracked the system’s evolution at several characteristic positions (A, B, C, D) to see the difference of  trajectories $\Delta \rm{T}= |\rm{T_{up}-T_{down}}|$, as given in Fig.~\ref{Fig3}(e).  As the transmission difference spectra have high discrimination, the intersection points of the trajectories under different microwave fields can be completely separated. In the region of hysteresis trajectories folding, the intersection points' nonlinear response to $\rm{\Omega}_{\rm{w}}$ offers a new approach to microwave field measurement.

When $\rm{\Omega_{w}}$ is small [$\rm{\Omega_{w}}/2 \pi \leq$ 2.85 ${\rm MHz}$], the hysteresis trajectories exhibit a clockwise evolution [Figs.~\ref{Fig3}(b)].The system is entirely in the particle-hole symmetry-broken phase and   possesses purely real eigenvalues $\lambda_i$ [Figs.~\ref{Fig3}(f)]. While for $\rm{\Omega_{w}/2} \pi \geq$ 8.1 ${\rm MHz}$, the hysteresis trajectories evolve in a counterclockwise direction, as shown in Figs.~\ref{Fig3}(d). This indicates that the system is entirely in the particle-hole symmetric phase, and the eigenvalues $\lambda_i$ are not purely real [Figs.~\ref{Fig3}(h)]. However, for 2.85 ${\rm MHz} < \rm{\Omega_{w}}/2 \pi < 8.1 \ {\rm MHz}$, the hysteresis trajectories are folded and intersect, as illustrated in Figs.~\ref{Fig3}(c). Within this parameter range, the system traverses both symmetric and symmetry-broken phases, undergoing a phase transition through the gap-closing point. In this case, the eigenvalues $\lambda_i$ change to purely real after passing through the EP point, as shown in Figs.~\ref{Fig3}(g). 

The role of the microwave field here is to build a population transfer channel between the Rydberg states $\ket{r_1}$ and $\ket{r_2}$, breaking the three-level EIT configuration. Thus, with a strong microwave field, the absorption of the probe field becomes dominant. This can also be revealed by the real parts of the eigenvalue \(E_3\) in Supplemental Material. The combination of population transfer and interaction-induced dissipation destroys the EIT coherence, thus leading to the decrease in transmission. Conversely, when $(\rm{\Omega_{p}}/2\pi)^2$ is reduced, the interaction-induced dissipation is alleviated, leading to a relative decrease in absorption within the system. These result in an opposite rotation direction by comparing with the scenario involving a weak microwave field. 

\subsection*{Phase boundaries versus scanning times}
In the experiment, we measured the phase diagrams within a scanning time range of [3 $\mu s$ - 20 $\mu s$] and obtained the phase boundaries by identifying the intersection points of the hysteresis trajectories, as shown in Fig.~\ref{Fig4}(a). The shape of the phase boundaries is dramatically dependent on the scanning time $\rm{T_s}$. At $\rm{T_s}\in$ [10 $\mu s$ - 20 $\mu s$], the system undergoes a hysteresis trajectory folding process analogous to that observed in Fig.~\ref{Fig3}(a), with overlapping trajectories to generate an intersection point $\mathrm{EP}_{\circlearrowleft}$ in the folding regime, see Fig.~\ref{Fig4}(b). 

\begin{figure*}
\centering
\includegraphics[width=1\linewidth]{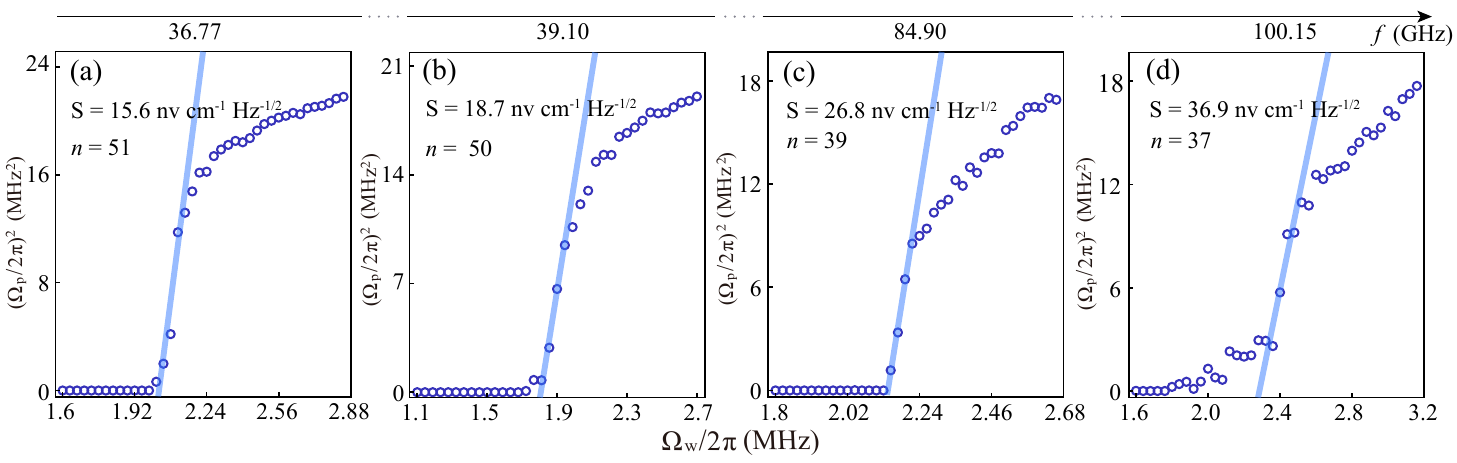}
\caption{\textbf{Measured phase boundaries at different principal quantum numbers.} (a)-(d) The phase boundaries measured at microwave frequencies of 36.77 GHz, 39.10 GHz, 84.90 GHz, and 100.15 GHz, respectively. The blue line represents the slope of the phase boundary. The measurements are enabled by tuning the principal quantum number n (or the corresponding microwave frequencies) to select different Rydberg transitions of the form $\ket{nD_{5/2}} \leftrightarrow \ket{(n-2)F_{7/2}}$ with $n = 51$, 50, 39, and 37 respectively. The sensitivity is estimated from the slope of the phase boundary near the critical region.}
\label{Fig6}
\end{figure*}

Furthermore, we observe more complex phase boundaries at scanning time $\rm{T_s}$ = 8 $\mu s$, where the hysteresis trajectory produces not only one intersection point $\mathrm{EP}_{\circlearrowleft}$ in Fig.~\ref{Fig4}(c). Another intersection point $\mathrm{EP}_{\circlearrowright}$ emerges due to the dynamic evolution of trajectory reversal. For a smaller scanning time $\rm{T_s}$ = 3 $\mu s$, the hysteresis trajectory is completely reversed. In this case, the intersection point $\mathrm{EP}_{\circlearrowleft}$ disappears, only leaving the intersection point $\mathrm{EP}_{\circlearrowright}$ [See Fig.~\ref{Fig4}(d)]. 

The underlying physical reason for this effect is that interaction-induced dissipation—arising from the accumulation of interactions over time—affects the coherence of Rydberg atoms and varies their response to the probe field. A relatively long (short) time interval $\rm{T_s}$ leads to a big (small) number of accumulated Rydberg atoms, resulting in strong (weak) interactions. In particular, the formation of trajectory intersection points arises from the competition between the system's transmission (gain) and absorption (dissipation) under varying microwave field conditions. When the scanning time is reduced, the interaction-induced dissipation diminishes, leading to a significant imbalance between the system's gain and dissipation. As a result, the folding region is observed to be reversed. As the microwave field varies, the hysteresis trajectory evolution transitions from counterclockwise to clockwise rotation [See Fig.~\ref{Fig4}(b) to Fig.~\ref{Fig4}(d)], and a completely different intersection point $\mathrm{EP}_{\circlearrowright}$ emerges during this transition. 

\subsection*{Sensitivities versus optical density}
To investigate how the system's sensitivity depends on Rydberg atomic interactions, we systematically varied the optical density (OD) calculated from probe absorption. OD reflects the absorption strength of the probe beam by the atomic ensemble, which is directly proportional to the atomic density and the effective length of the atomic cloud~\cite{zhang2012dark}. As shown in Figs.~\ref{Fig5}(a)-(c), the relatively large OD provides strong interactions between Rydberg atoms, which leads to the larger slope for the phase boundary, defined as $k=d\rm{(\Omega_p/2\pi)^2}/$$d$$\rm{(\Omega_w/2\pi)}$. This relationship enables us to quantitatively characterize how the phase boundary slope varies with OD, revealing the system's susceptibility at the criticality. In Fig.~\ref{Fig5}(d), we find that the phase boundary slope exhibits exponential growth with increasing OD. Thus, a higher sensitivity could be achieved at a larger OD.

In the experiment, the sensitivity is determined by both the variances in the transmitted signal near the intersection point and the optical intensity noise. The phase boundaries are obtained from a single measurement, where the measured transmission spectra of the Up and Down processes near the intersection point are shown in Figs.~\ref{Fig5}(e)-(g). We use $\sigma$ to quantify the measurement precision limit; here $\sigma$ also characterizes the spectral overlap range of the Up and Down transmission lines. The black data points in Figs.~\ref{Fig5}(e)-(g) represent the transmission difference spectrum $\Delta \rm{T}$. 

By analyzing the displacement $\delta\rm{(\Omega_p/2\pi)^2}$ and average uncertainty $\bar{\sigma}$ of the intersection points in the critical region, we estimate the sensitivity expression: $\rm{S}=[{\bar{\sigma}}/{\delta\rm{(\Omega_p/2\pi)^2}} ] \rm{\Delta E_w}\sqrt{T}$, where $\Delta \rm{E_w}$ corresponds single step size of changing microwave intensity $\rm{E_w}$ during measurement time T. The results of the sensitivity are presented in Fig.~\ref{Fig5}(h), where the sensitivity uncertainty originates from the fluctuation of the fitted uncertainty $\sigma$ at the intersection point. It can be seen that the sensitivity gradually increases with the increase of OD, and the maximum sensitivity of the system reaches 1.6(5) $\rm{nV}cm^{–1}Hz^{–1/2}$. The power sensitivity is calculated as $\rm{P} = 10log_{10}(\varepsilon_0 c \text{S}^2 A/10^{-3}) \approx -208.4 \pm 2.7~\text{dBm} ~\mathrm{Hz^{-1}}$\cite{fancher2021rydberg}. Here, $\varepsilon_0$ is the vacuum permittivity, $c$ is the speed of light, $\rm{S}$ denotes the electric field sensitivity, and A is the effective detection area.

To verify the quantum-enhancement in this protocol enabled by the interaction between Rydberg atoms, we compared the measurement sensitivity with the standard quantum limit (SQL) as a function of the OD, as shown by the gray curve in Fig. \ref{Fig5} (h). The SQL is expressed as $\text{S}_{\text{SQL}} =\frac{ \textit{h}}{\mu_ \text{w} \sqrt{N\text{T}}}$\cite{bussey2022quantum}, where $h$ is Planck constant, $\mu_ \text{w}$ is the microwave transition dipole moment, and T is the measurement time. The atom number $N$ is estimated from the OD via the relation  $N= {m\times\mathrm{OD}}/{a_0}$ \cite{zhang2012dark}, where $a_0$ is the resonant absorption cross section, and \(m\) is the cross-sectional area of the beam. It can be seen that at low OD (corresponding to the weakly interacting regime), the measured sensitivity remains above the SQL. As OD increases (reflecting stronger interactions), the sensitivity is progressively enhanced and eventually surpasses the SQL, demonstrating the superior performance in the strong-interaction regime.

\subsection*{Measured phase boundaries at different principal quantum numbers}
One advantage of Rydberg atomic systems for microwave sensing is their frequency tunability\cite{simons2016simultaneous}, achieved by adjusting the principal quantum number $n$ to match different microwave transition. To experimentally validate the applicability of the sensing scheme across a broad frequency range, we systematically measured the phase boundaries at several distinct microwave frequencies from 36.77 GHz $\sim$ 100.15 GHz.  The intersection points of hysteresis trajectories folding exhibit nonlinear dynamics across all tested frequencies. We estimated the corresponding sensitivity by analyzing the slope of phase boundaries in the vicinity of the critical regime, as shown in Figs. \ref{Fig6} (a)-(d). 

In this process, a smaller $n$ corresponds to a smaller transition dipole moment [for example, $\mu_{\mathrm{w}} = 1292 ea_0 $ for $n$ = 51, and $\mu_{\mathrm{w}} = 651 ea_0 $ for $n$ = 37\cite{vsibalic2017arc}], inducing lower sensitivity in the sensing process. In addition, the interaction between Rydberg atoms becomes weak when $n$ decreases, consequently reducing the interaction-induced non-Hermitian effect. While the sensitivity decreases with $n$, our system maintains its capability for ultra-broadband microwave frequency detection, offering unparalleled versatility across an extensive spectral range.

\section*{Discussions}

This work establishes a paradigm for quantum-enhanced sensing by many-body effects in cold atomic ensembles with the geometric trajectories. Unlike single-particle or weakly interacting systems, many-body interactions in Rydberg atom systems offer a unique platform to study critical effects on external electric fields. These systems exhibit collective dynamical responses to external fields, where perturbations propagate coherently across the ensemble rather than affecting individual atoms independently. Such collective dynamics render the ensemble exceptionally sensitive to external fields (e.g., microwave field and probe field), making them ideal for two-parameter quantum sensing applications. Importantly, the tunability of interactions allows precise control over the system’s evolution, enabling measurements beyond conventional methods. This interplay of cooperative many-body behavior and tailored quantum evolution distinguishes the approach from classical sensing paradigms, establishing a robust framework for high-precision metrology.

In the experiment, a larger OD increases the interaction strength between Rydberg atoms, thereby improving the sensing sensitivity. Similarly, increasing the principal quantum number further strengthens these interactions, making the system more responsive to external perturbations. However, higher principal quantum numbers require lower probe field intensities to produce sharper phase boundaries. Reducing the probe field intensity leads to larger fluctuations near the intersection points, which degrades measurement sensitivity. Thus, an optimal balance between the principal quantum number and probe field intensity is essential to maximize sensitivity while minimizing noise. 

In our scheme, the nonlinear variation of the hysteresis loops' intersection point with the microwave field provides a resource for enhanced sensing, as the interaction-induced dissipation is the key mechanism behind the trajectory flipping in the system. By adjusting the scanning time, OD and the principal quantum number, the Rydberg atom interactions can be tuned to induce distinct phase boundaries. Under optimized experimental parameters, tracking the response of the intersection point to external fields achieves a sensitivity of 1.6(5) nV/cm$^{–1}$Hz$^{–1/2}$. This result demonstrates the Rydberg many-body system’s significant advantage in precision measurements.

\section*{Methods}

\begin{figure*}
\centering
\includegraphics[width=0.9\linewidth]{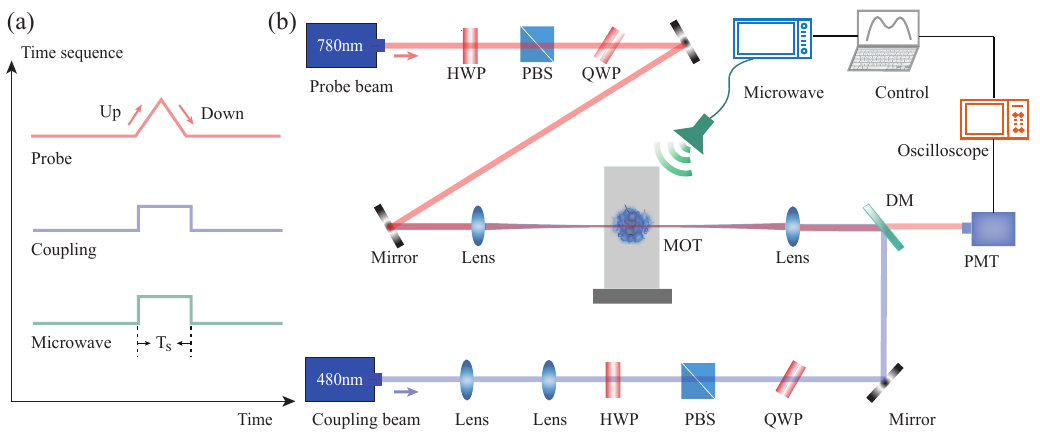}
\caption{\textbf{Experimental setup and time sequence.} (a) The partial time sequence of the experiment, in which the probe beam is modulated with a triangular wave. (b) The diagram of the experimental setup. The probe beam is incident opposite to the coupling beam and focused in cold $^{85}$Rb atoms. A horn radiates microwave electric field on atoms. The following optical components are briefly described in the figure: half-wave plate (HWP), quarter-wave plate (QWP), polarizing beam splitter, (PBS), magneto-optical trap (MOT), dichroic mirror (DM), and photomultiplier tube (PMT).}
\label{FigSS4}
\end{figure*}

\textbf{Experimental setup} To study hysteresis trajectories folding, we prepare a cold ensemble of $^{85}$Rb atoms trapped in a three-dimensional magneto-optic trap [See Fig.~\ref{FigSS4}]. The atomic ensemble is prepared in the ground state $\ket{g} = \ket{5S_{1/2}, F = 3}$ by an optical pumping process. In our experiment, we shield the magneto-optical trap with a double-layer magnetic shielding system. This setup effectively shields the system from external magnetic fields and lowers the internal magnetic field to below 10 mGauss. This configuration can avoid the dephasing from the earth's magnetic field.

We used a two-photon transition scheme to excite $^{85}$Rb atoms from the ground state to the Rydberg state. The probe field drives the atoms from the ground state $\ket{g}$ to the intermediate excited state $\ket{e} = \ket{5P_{3/2}, F = 4}$, and then the coupling field  drives the transition from $|e\rangle$ to the Rydberg state $\ket{r_1} = \ket{42D_{5/2}}$. We further used a microwave electric field to drive the transition between two different Rydberg states, $\ket{r_1}$ and $\ket{r_2}=|41F_{7/2}\rangle$, with a frequency of $2\pi\times$31.889 GHz. In the experiment, the detuning values for the coupling, probe, and microwave field were set to zero, which is consistent with the conditions used in the theoretical calculations. The probe field and coupling field are focused into the cold atomic ensemble, and the transmittance of the EIT is obtained by detecting the intensity of the probe field via a photo-multiplier tube. 

\textbf{Automated measurement} We loaded a triangular waveform that was generated using a signal generator (RIGOL DG4102) onto the acousto-optic modulator. The functional form of the triangular wave signal is given by $V(\tau) = 6 \left( 1/2 -  \left| \tau /\rm{T_{s}} - 1/2 \right| \right)$, where $\rm{T_{s}}$ represents the period time of one scanning cycle and $0 \leq \tau \leq \rm{T_{s}}$; the time sequence is shown in Fig.~\ref{FigSS4}(a). By this way, we produce the process of increasing $(\rm{\Omega _{p} / 2 \pi )^2}$ from 0 ${\rm MHz}^2$ to 40 ${\rm MHz}^2$ (Up process) and the process of decreasing $(\rm{\Omega _{p} / 2 \pi )^2}$ from 40 ${\rm MHz}^2$ to 0 ${\rm MHz}^2$(Down process). Additionally, a computer-controlled microwave RF source generates the signal, while an oscilloscope automatically records and saves the measurement data.

\begin{figure*}
\centering
\includegraphics[width=0.95\linewidth]{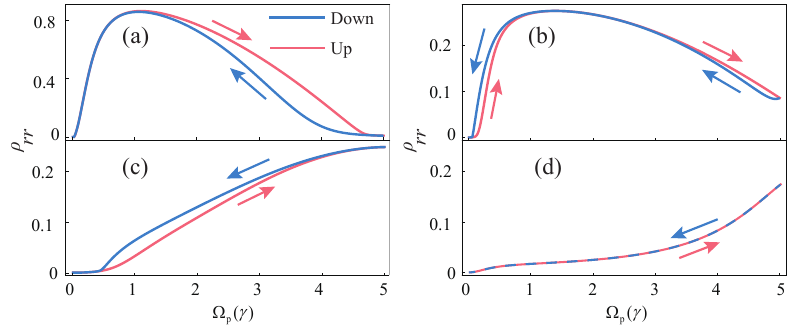}
\caption{\textbf{The population of $\ket{r}$ state simulated by the master equation.} 
The population of hysteresis trajectories $\rho_{rr}$ under microwave Rabi frequency $\rm{\Omega_{w}}$ = 0.1$\gamma$ (${\rm a}$),  $\rm{\Omega_{w}}$ = 2.2$\gamma$ (${\rm b}$), and $\rm{\Omega_{w}}$ = 5.5$\gamma$ (${\rm c}$), respectively. The other parameters are $\{ V,\rm{\Omega_c}, \gamma_1, \Gamma, \gamma_3\}=\{-800,0.3,0.5,1,1\} $, the parameter $\rm{\Omega_{p}}$ varies within the range [0 - 5$\gamma$], and all energy detuning are 0. The dashed lines in (${\rm d}$) represents the hysteresis trajectory under the condition $V=0$. The pink and blue arrows represents the positive and negative scanning $\rm{\Omega_{p}}$, respectively.}
\label{Fig7}
\end{figure*}

\textbf{Calibration of the microwave fields} In our experiment, the microwave electric field was generated by a broadband RF signal generator and a horn antenna for frequencies up to 40 GHz. Additionally, high-frequency microwave signals (e.g., 84.90 GHz and 100.15 GHz) were generated via a frequency-sextupling terahertz link. The microwave electric field amplitude $\rm{ E_{\mathrm{w}}}$ was calibrated via Autler–Townes (AT) splitting observed in the EIT spectrum, see more detail in the Supplemental Material. A resonant microwave field was applied to drive the Rydberg transition, inducing Autler-Townes splitting with a frequency separation $\Delta w$, which is directly related to the microwave Rabi frequency by $\rm{\Omega_w}$ = $\Delta w$. The electric field amplitude $\rm{E_{\mathrm{w}}}$ is given by 
$\rm{ E_{\mathrm{w}} = \frac{\hbar \Omega_w}{\mu_{\mathrm{w}}} \cdot \frac{\lambda_p}{\lambda_c}}$, where  $\mu_{\mathrm{w}}$ denotes the dipole moment of the microwave transition, $\hbar$ is the reduced Planck constant, and \(\lambda_p\), \(\lambda_c\) are the wavelengths of the probe and coupling beams, respectively.

\textbf{Lindblad master equation} In the cold atomic ensemble, atoms excited to the Rydberg state experience a field arising from their many-body interactions with other atoms. We can effectively treat this interaction using the mean-field approximation and solve the dynamical evolution of the density matrix elements via the Lindblad master equation $\dot{\rho}=-i[\hat{H},\rho]+L[\rho]$. Where $L[\rho] = \sqrt{\gamma_1}\ket{g} \bra{e}+\sqrt{\gamma_2}\ket{e} \bra{r_1}+\sqrt{\gamma_3}\ket{r_1} \bra{r_2}$, denotes the spontaneous emission rate from the highly excited state to the lower excited state. Below are the time-dependent density matrix elements obtained by solving the Lindblad master equation.
\begin{equation}
\begin{aligned}
&\dot\rho_{g g} = \gamma_1 \rho_{ee} + \frac{i}{2} \rm{\Omega_{p}} (-\rho_{eg} + \rho_{ge}), \\
&\dot\rho_{g e} = \frac{i}{2} \Big( i \gamma_1 \rho_{ge} - 2 \Delta_1 \rho_{ge} + \rm{\Omega_{c}}\rho_{g r_1}+\rm{\Omega_{p}} (\rho_{gg} - \rho_{ee}) \Big), \\
&\dot\rho_{g r_1} = \frac{i}{2} \Big( i \gamma_2 \rho_{g r_1} - 2 (\Delta_1 + \Delta_2) \rho_{g r_1} +\rm{\Omega_{c}}\rho_{ge}- \rm{\Omega_{p}} \rho_{e r_1} \\
&\quad\quad\quad+ \rm{\Omega_{w}} \rho_{g r_2} \Big), \\
&\dot\rho_{g r_2} = \frac{1}{2} \Big( -\gamma_3 \rho_{g r_2} - 2 i (\Delta_1 + \Delta_2 + \Delta_3) \rho_{g r_2}\\
&\quad\quad\quad-i \rm{\Omega_{p}} \rho_{e r_2} + i \rm{\Omega_{w}} \rho_{g r_1} \Big),\nonumber
\end{aligned}
\end{equation}
\begin{equation}
\begin{aligned}
&\dot\rho_{e e} = \frac{i}{2} \Big( 2 i \gamma_1 \rho_{ee} - 2 i \gamma_2 \rho_{r_1 r_1} + \rm{\Omega_{c}}(\rho_{e r_1} - \rho_{r_1 e})\\ 
&\quad\quad\quad+ \rm{\Omega_{p}} (\rho_{eg} - \rho_{ge}) \Big), \\
&\dot\rho_{e r_1}= -\frac{i}{2} \Big( -i (\gamma_1 + \gamma_2) \rho_{e r_1} + 2 \Delta_2 \rho_{e r_1} + \rm{\Omega_{c}}(\rho_{r_1 r_1} - \rho_{ee})\\
&\quad\quad\quad+ \rm{\Omega_{p}} \rho_{g r_1} - \rm{\Omega_{w}} \rho_{e r_2} \Big),\\ 
&\dot\rho_{e r_2}= -\frac{i}{2} \Big( -i (\gamma_1 + \gamma_3) \rho_{e r_2} + 2 (\Delta_2 + \Delta_3) \rho_{e r_2}\\
&\quad\quad\quad+ \rm{\Omega_{c}}\rho_{r_1 r_2} + \rm{\Omega_{p}} \rho_{g r_2} - \rm{\Omega_{w}} \rho_{e r_1} \Big), \\
&\dot\rho_{r_1 r_1}= -\frac{i}{2} \Big( -2 i \gamma_2 \rho_{r_1 r_1} + 2 i \gamma_3 \rho_{r_2 r_2} + \rm{\Omega_{c}}(\rho_{e r_1} - \rho_{r_1 e})\\
&\quad\quad\quad- \rm{\Omega_{w}} (\rho_{r_1 r_2} - \rho_{r_2 r_1}) \Big),\\
&\dot\rho_{r_1 r_2}= -\frac{i}{2} \Big( -i (\gamma_2 + \gamma_3) \rho_{r_1 r_2} + 2 \Delta_3 \rho_{r_1 r_2} + \rm{\Omega_{c}}\rho_{e r_2} \\
&\quad\quad\quad+ \rm{\Omega_{w}} (\rho_{r_2 r_2} - \rho_{r_1 r_1}) \Big), \\
&\dot\rho_{r_2 r_2} = -\gamma_3 \rho_{r_2 r_2} - \frac{1}{2} i (\rho_{r_1 r_2} - \rho_{r_2 r_1}) \rm{\Omega_{w}}.
\end{aligned}
\end{equation}
The other expressions satisfy $\rho_{e g}=\rho_{ge}^\dagger$, $\rho_{r_1 g}=\rho_{gr_1}^\dagger$, $\rho_{r_2 g}=\rho_{gr_2}^\dagger$, $\rho_{r_1 e}=\rho_{er_1}^\dagger$, $\rho_{r_2 e}=\rho_{er_2}^\dagger$, $\rho_{r_2 r_1}=\rho_{r_1 r_2}^\dagger$, and the Rydberg population is normalized by the expression $1 = \rho_{ee} + \rho_{gg} + \rho_{r_1 r_1} + \rho_{r_2 r_2}$. Under the mean field approximation $\gamma_2\rightarrow\Gamma+V\rho_{r_1 r_1} $, the dissipation induced by the interaction between Rydberg states is proportional to the population of Rydberg states $\ket{r_1}$. From the lindblad master equation, we can obtain the population of Rydberg states $\ket{r_1}$, written as
\begin{equation}
\begin{aligned}
\rho_{r_1 r_1}= &\frac{1}{i2\gamma_2}\Big( \frac{i}{2} \dot {\rho}_{r_1 r_1} + 2 i \gamma_3 \rho_{r_2 r_2} + \rm{\Omega_{c}}(\rho_{e r_1} - \rho_{r_1 e})\\
&- \rm{\Omega_{w}} (\rho_{r_1 r_2} - \rho_{r_2 r_1}) \Big).
\end{aligned}
\end{equation}
By setting the initial moment $\rho_{gr_1} [0]=0$, $\rho_{er_1} [0]=0$, $\rho_{gr_2}[0]=0$, $\rho_{r_1g}[0]=0$, $\rho_{r_1e} [0]=0$ and $\rho_{r_2g}[0]=0$, we solve the time-dependent evolution of the system employing the Lindblad master equation as follows
\begin{equation}
\begin{aligned}
\frac{{\partial \rho (t)}}{{\partial t}} =  - i[H(t),\rho (t)] + \sum\limits_i {({L_i}} \rho (t)L_i^\dagger  - \frac{1}{2}\{ L_i^\dagger {L_i},\rho (t)\}).  
\end{aligned}
\end{equation}

We investigate the trajectory loop of the population $\rho_{r_1r_1}$ by performing both forward (Up) and backward (Down) scans of the probe field Rabi frequency $\rm{\Omega_{p}}$, as depicted in Fig.~\ref{Fig7}(b). The interactions between Rydberg atoms break the symmetry of the system and give rise to non-closed hysteresis trajectories \cite{zhang2025exceptional}. As the microwave Rabi frequency $\rm{\Omega_{w}}$ increases, the trajectories of $\rho_{r_1r_1}$ exhibit different patterns (Figs.~\ref{Fig7}(a), (b) and (c)), corresponding to distinct trajectories of the system. Additionally, we also present the evolution curve of the Rydberg state population as a function of $\rm{\Omega_{p}}$ when the interaction between Rydberg atoms is zero.

\hspace*{\fill}

\section*{Data Availability}
The data generated in this study have been deposited in the Zenodo
database (\href{https://zenodo.org/records/17451657}{https://zenodo.org/records/17451657}).\\

\section*{Code availability}
The custom codes used to produce the results presented in this paper are available from the corresponding authors upon request.

\section*{Acknowledgements}
We acknowledge funding from the National Key R and D Program of China (Grant No. 2022YFA1404002), the National Natural Science Foundation of China (Grant Nos. T2495253, 62435018), and the Major Science and Technology Projects in Anhui Province (Grant No. 202203a13010001).

\section*{Author contributions statement}
D.-S.D., L.-H.Z. and B.L. conceived the idea. J.Z, Y.J.W, and Z.Y.Z conducted the physical experiments. D.-S.D. and Y.J.W developed the theoretical model. The manuscript was written by D.-S.D, J.Z., Y.J.W., and Z.Y.Z. The research was supervised by D.-S.D. All authors contributed to discussions regarding the results and the analysis contained in the manuscript.

\section*{Competing interests}
The authors declare no competing interests.

\bibliography{ref}

\maketitle

\onecolumngrid

\appendix
\newpage

\begin{center}
\textbf{Supplemental Material for \lq \lq Quantum enhanced metrology based on flipping trajectory of cold Rydberg gases"}
\end{center}

\begin{center}
{Ya-Jun Wang$^{1,2,\textcolor{blue}{*}}$, Jun Zhang$^{1,2,\textcolor{blue}{*}}$, Zheng-Yuan Zhang$^{1,2,\textcolor{blue}{*}}$, Shi-Yao Shao$^{1,2}$, Qing Li$^{1,2}$, Han-Chao Chen$^{1,2}$, Yu Ma$^{1,2}$\\ Tian-Yu Han$^{1,2}$, Qi-Feng Wang$^{1,2}$, Jia-Dou Nan$^{1,2}$, Yi-Ming Yin$^{1,2}$, Dong-Yang Zhu$^{1,2}$, Qiao-Qiao Fang$^{1,2}$, Chao\\ Yu$^{1,2}$, Xin Liu$^{1,2}$, Guang-Can Guo$^{1,2}$, Bang Liu$^{1,2,\textcolor{blue}{\ddagger}}$, Li-Hua Zhang$^{1,2,\textcolor{blue}{\S}}$, Dong-Sheng Ding$^{1,2,\textcolor{blue}{\dagger}}$, and Bao-Sen Shi$^{1,2}$}
\end{center}
\begin{center}
\textit{$^1$Laboratory of Quantum Information, University of Science and Technology of China, Hefei, Anhui 230026, China.} \\
\textit{$^2$Synergetic Innovation Center of Quantum Information and Quantum Physics, \\University of Science and Technology of China, Hefei, Anhui 230026, China.} \\
\end{center}

\symbolfootnote[1]{Y.J.W, J.Z, and Z.Y.Z contribute equally to this work.}

\symbolfootnote[3]{lb2016wu@ustc.edu.cn}
\symbolfootnote[4]{zlhphys@ustc.edu.cn}
\symbolfootnote[2]{dds@ustc.edu.cn}

The Supplemental Materials contain the following content to supplement the main text. Firstly, the particle-hole symmetry breaking of the system is analyzed in the theoretical model section. Secondly, the sensitivity advantage of our microwave field detection scheme over conventional heterodyne detection is analyzed.
\begin{center}
\textbf{Theoretical model}
\end{center}

\begin{figure*}[b]
\centering
\includegraphics[width=1\linewidth]{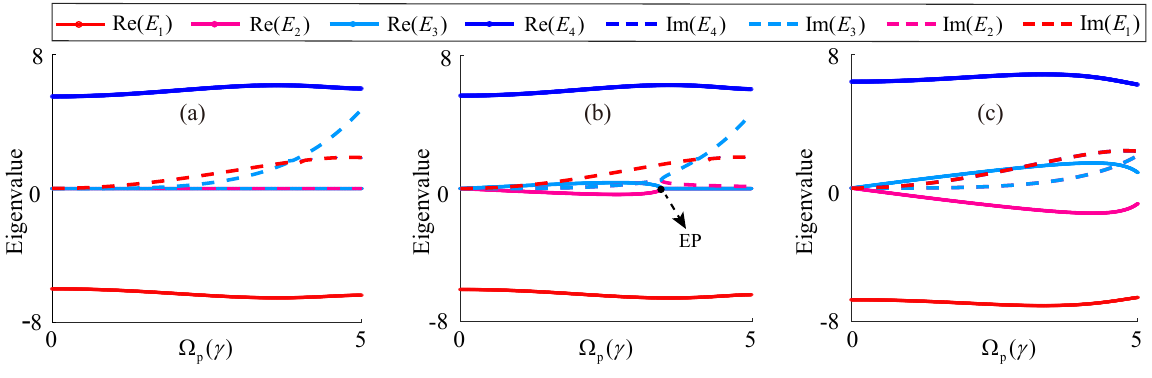}
\caption{\textbf{The Energy spectrum.} The real (solid line) and imaginary (dotted line) parts of the system's eigenvalues are plotted as a function of $\rm{\rm{\Omega_{p}}}$ with $\rm{\Omega_c}$ = 6$\gamma$ and $V_{eff}= V \rho_{r_1r_1}$ $(\rho_{r_1r_1}\propto\rm{\Omega_{p}}^2)$ in cases of (a) $\rm{\Omega_{w}}$ = 0.01$\gamma$, (b) $\rm{\Omega_{w}}$ = $\gamma$, (c) $\rm{\Omega_{w}}$ = 3$\gamma$. }
\label{FigS1}
\end{figure*}

Highly excited Rydberg atoms exhibit strong long-range interactions due to their large electric dipole moments. These strong interactions between Rydberg atoms can cause spatially dependent interaction-induced dissipation \cite{busche2017contactless,maxwell2013storage,dudin2012strongly,bariani2012dephasing}. In our system, the anisotropic dipole-dipole interactions induce phase shifts that depend on both distance and direction. The resulting inhomogeneous phase shifts lead to the population-dependent dissipation in the Rydberg atom system and increase the total relaxation rate \cite{ding2019Phase,de2016intrinsic,qian2014dissipation}. These effects collectively give rise to an effective dissipation rate $ \gamma_{\rm eff}$ for the Rydberg state $|r_1\rangle$. The effective non-Hermitian Hamiltonian is expressed as:
\begin{equation}
\hat{H}= \hbar /2\left( {\begin{array}{*{20}{c}}
0&{\rm{\rm{\Omega_{p}}}}&0&0\\
{\rm{\rm{\Omega_{p}}}}&{ - 2\Delta_1}&{\rm{\Omega_c}}&0\\
0&{\rm{\Omega_c}}&{ - 2\Delta_{r_1}-iV_{eff}}&{\rm{\rm{\Omega_{w}}}}\\
0&0&{\rm{\rm{\Omega_{w}}}}&{ - 2 \Delta_{r_2} }
\end{array}} \right) ,
\end{equation}
where $\Delta _{r_1}=\Delta _1+\Delta _2$ and $\Delta _{r_2}=\Delta _1+\Delta _2+\Delta _3$. The term $V_{eff} ={{V\rho_{rr}}}/{2}$, derived from  mean-field treatment, represents interaction-induced dissipation and effectively broadens the energy level $|r_1\rangle$ \cite{zhang2025exceptional}. 

This non-Hermitian Hamiltonian has particle-hole symmetry
\begin{align}
\mathcal CH^*{\mathcal C^\dag } =  - H,
\end{align}
where $\mathcal C=\mathcal I\otimes {\sigma _z}\mathcal K$, $\mathcal I$ is the identity matrix, ${\sigma _z}$ is the Pauli matrix acting, and $\mathcal K$ is the complex conjugation acting on the wave function of the whole system. The operation of the Pauli matrix ${\sigma _z}$ imposes opposite phases on the Rydberg states $\ket{r_1}$ and $\ket{r_2}$, and the complex conjugation operator $\mathcal K$ reverses the direction of time evolution ($i\rightarrow-i$). When the microwave Rabi frequency $\rm{\Omega_{w}}$ varies, the system's particle-hole symmetry is spontaneously broken, leading to a gap-closing phenomenon. 

By setting the detunings $\Delta_{1} = \Delta_{2} = \Delta_{3} = 0$, we calculate the eigenvalues of the above non-Hermitian Hamiltonian, denoted as $E_{1} \sim E_{4}$. Then, we obtain the real and imaginary parts of the system's eigenvalues as a function of $\rm{\Omega_{p}}$, as depicted in Fig.~\ref{FigS1}. One sees that the overall properties of the system are mainly affected by two energy levels near zero energy. In our parameter space, when the microwave Rabi frequency is $\rm{\Omega_{w}}$ = 0.01$\gamma$ in Fig.~\ref{FigS1}(a), the relationship between these levels is ${\rm{Re}}(E_{2})=-{\rm{Re}}(E_{3})$ and ${\rm{Im}}(E_{2}) \ne {\rm{Im}}(E_{3})$, indicating that the particle-hole symmetry of the system is completely broken. At $\rm{\Omega_{w}}$ = $\gamma$ in Fig.~\ref{FigS1}(b), it is found that the particle-hole symmetry of the system is spontaneously broken, and the broken defect is at the `gap-closing' point. However, when the microwave Rabi frequency is $\rm{\Omega_{w}}$ = 3$\gamma$ in Fig.~\ref{FigS1}(c), the system maintains particle-hole symmetry, and the two eigenvalues have a relationship of $E_2=-{E_3}^*$ near the zero energy.

Next, we depict the evolution of eigenvalues in the energy spectrum to reveal the physical process. When we set $\rm{\Omega_{w}} = $0.01$\gamma$, the pure non-zero imaginary eigenvalues for \(E_3\) (real parts are zero) are observed in Fig.~\ref{FigS1}(a), highlighting the non-Hermitian property of the system. The imaginary parts create an additional dissipation, thus inducing a dynamical hysteresis trajectory in the experiment, as discussed in the main text. In Fig.~\ref{FigS1}(c), we can see that the eigenvalues \(E_2\) and \(E_3\) contain both real and imaginary components. As illustrated in Figs.~\ref{FigS1}(a) and (c), the disparity in the real parts of \(E_2\) and \(E_3\) gives rise to two distinctly different hysteresis trajectories in the experiment. In Fig.~\ref{FigS1}(a), the purely imaginary components contribute to a hysteresis trajectory in a clockwise direction.

Conversely, under the influence of a strong microwave field, the energy shift of the state \(\left| r \right\rangle\) becomes significant, leading to the system's absorption. This results in an opposite surrounding direction for the trajectory. Furthermore, in Fig.~\ref{FigS1}(b), the eigenvalues $E_2$ and $E_3$ display a `gap-closing' spectral behavior. As the driving parameter $\rm{\rm{\Omega_{p}}}$ is varied, the spectrum undergoes a transition from possessing real components to becoming entirely imaginary at the `gap-closing' point. This results in the formation of a hysteresis loop transitioning from counterclockwise to clockwise.

\begin{figure*}[b]
\centering
\includegraphics[width=1\linewidth]{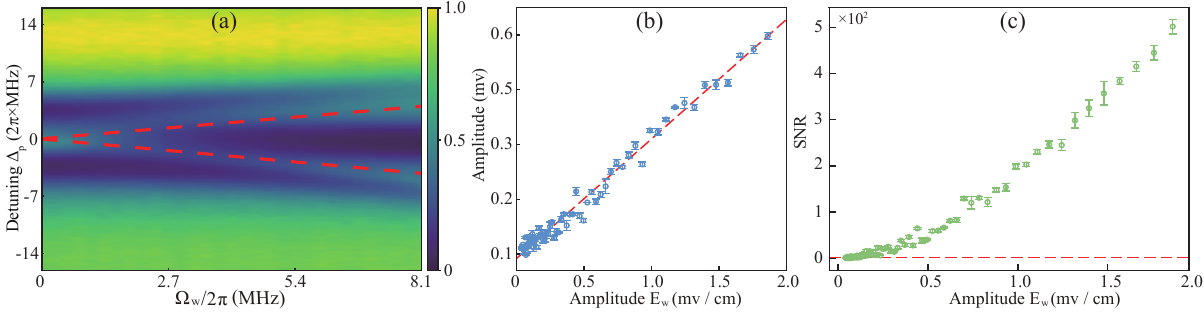}
\caption{\textbf{Heterodyne detection performance.} (a)Measured phase diagram of the Autler–Townes spectrum versus microwave Rabi frequency, and the red dashed line represents the linear fit of the peak. (b) The measured oscillation amplitude as a function of microwave field amplitude $\text{E}_{\text{mw}}$. The red dashed line represents the fitted linear response curve of the amplitude. (c) The signal-to-noise ratio (SNR) extracted from the FFT analysis as a function of $\text{E}_{\text{mw}}$. The red dashed line in indicates the threshold where $\text{SNR}=2$, which is used to estimate the system sensitivity, yielding a value of S = 2300 $\text{nV cm}^{–1}$ Hz$^{–1/2}$.}
\label{FigS2}
\end{figure*}

\begin{center}
\textbf{Comparison between trajectory folding and heterodyne detection methods}
\end{center}

To demonstrate the superiority of our proposed scheme in terms of measurement sensitivity, we present a comparative analysis with the conventional heterodyne measurement method. Heterodyne detection converts weak microwave fields into detectable optical signals, making it an effective measurement approach in Rydberg atomic systems \cite{jing2020atomic}. In the experiment, the local microwave field resonantly drives the Rydberg-state transition to form microwave-dressed states, while the signal microwave field introduces a beat-note perturbation. 

As shown in Fig.~\ref{FigS2}(a), we first measure the Autler–Townes (AT) splitting spectra as a function of the microwave Rabi frequency to calibrate the microwave field strength. The field amplitude $\mathrm{E_{w}}$ is given by 
$\mathrm{E_{w}} = \frac{\hbar \Omega_{\mathrm{w}}}{\mu_{\mathrm{w}}} \cdot \frac{\lambda_{\mathrm{p}}}{\lambda_{\mathrm{c}}}$, where $\mu_{\mathrm{w}}$ is the dipole matrix element between the relevant Rydberg states, calculated using the Alkali Rydberg Calculator (ARC) package~\cite{vsibalic2017arc}. The red dashed line in Fig.~\ref{FigS2}(a) represents the linear fit of the AT peak positions, confirming the proportional relationship between the splitting and the Rabi frequency. After calibration, the power of the local microwave field is adjusted to place the EIT spectrum at its most sensitive point, and the transmitted probe signal is recorded. 

Figure.~\ref{FigS2}(b) presents the measured oscillation amplitude versus the microwave field amplitude $\text{E}_{\text{mw}}$, showing a linear response relationship. The data points align well with the red dashed linear fit, indicating a stable and predictable response of the system to microwave fields. Figure.~\ref{FigS2}(c) illustrates the signal-to-noise ratio (SNR) obtained through fast Fourier transform (FFT) analysis as a function of microwave field amplitude. The system sensitivity is estimated to be 2300 $\text{nV cm}^{–1}\text{ Hz}^{–1/2}$, with the threshold SNR=2 (red dashed line) serving as a reference for sensitivity estimation. This relatively lower heterodyne sensitivity mainly arises from the experimental configuration. A tightly focused probe beam with a waist of $\omega_p \approx 20~\mu$m and a reduced optical depth (OD) were used to suppress interactions between Rydberg atoms that could distort the beat-note signal, which together reduced the effective number of atoms contributing to the heterodyne signal. Moreover, a longer averaging time was required to obtain a stable beat-note signal.

However, the system demonstrated a sensitivity of 1.6(5) $\rm{nV cm^{-1}~Hz^{-1/2}}$ in hysteresis trajectory measurements. This indicates that the scheme we proposed for measuring the sensitivity of cold Rydberg atomic systems by tracking the intersection points (gap-closing points) of hysteresis trajectories has significant advantages.


\end{document}